\documentclass[lengthestimate,tighten,amssymb,prb,aps,twocolumn,floatfix,tightenlines]{revtex4}

\usepackage{graphicx,subfigure}
\usepackage{amsfonts}
\usepackage{amsmath}
\usepackage{lettrine}
\usepackage{type1cm}
\usepackage{setspace}

\newcommand{\ket}[1]{\left | #1 \right \rangle}
\newcommand{\bra}[1]{\left \langle #1 \right |}
\def \kron{\raisebox{0.3pt}{\ensuremath{\:\otimes\:}}} 
\begin{document}
\title{Generating,~manipulating~and~measuring~entanglement~and~mixture~with a reconfigurable photonic circuit}

\author{P. J. Shadbolt}
\author{M. R. Verde}
\author{A. Peruzzo}
\author{A. Politi}
\author{A. Laing}
\author{M. Lobino}
\author{J. C. F. Matthews}
\author{M. G. Thompson}
\author{J. L. O'Brien}
\email{Jeremy.OBrien@bristol.ac.uk}
\affiliation{Centre for Quantum Photonics, H. H. Wills Physics Laboratory \& Department of Electrical and Electronic Engineering, University of Bristol, Merchant Venturers Building, Woodland Road, Bristol, BS8 1UB, UK}

\begin{abstract}
Entanglement is the quintessential quantum mechanical phenomenon understood to lie at the heart of future quantum technologies and the subject of fundamental scientific investigations. Mixture, resulting from noise, is often an unwanted result of interaction with an environment, but is also of fundamental interest, and is proposed to play a role in some biological processes.
Here we report an integrated waveguide device that can generate and completely characterize pure two-photon states with any amount of entanglement and arbitrary single-photon states with any amount of mixture. The device consists of a reconfigurable integrated quantum photonic circuit with eight voltage controlled phase shifters. We demonstrate that for thousands of randomly chosen configurations the device performs with high fidelity. We generate maximally and non-maximally entangled states, violate a Bell-type inequality with a continuum of partially entangled states, and demonstrate generation of arbitrary one-qubit mixed states. 
\end{abstract}

\maketitle

\lettrine[lines=3,slope=0pt,nindent=4pt, lraise=0.1, loversize=-0.1]{Q}{uantum} mechanics is known to allow fundamentally new modes of information processing\cite{niels-a, deuts-potrsolamaps-400-97}, simulation\cite{feynm-ijotp-21-467, lloyd-s-273-1073}, and communication \cite{gisin-np-1-165} as well as enhanced precision of measurement and sensing \cite{giova-s-306-1330, dowli-cp-49-125}. Single photons provide a particularly promising physical system with which to develop such quantum technologies\cite{o'bri-np-3-687}---due to their low noise, high speed transmission and ease of manipulation at the single photon level---and have long been a leading approach to exploring fundamental quantum science. The ability to precisely prepare, control and measure multi-photon states therefore holds considerable scientific and technological interest. 

Recently it has been shown that it is possible to miniaturize quantum optical circuits using optical fibre \cite{clark-pra-79-30303,hall-prl-106-53901} and  integrated waveguide chips \cite{polit-s-320-646, matth-np-3-346, smith-oe17-13516, polit-s-325-1221, sanso-prl-105-200503, laing-apl-97-211109, peruzz-nc-2-224}. Monolithic waveguide circuits are inherently stable and can be many orders of magnitude smaller than their bulk optical equivalents, enabling the fabrication of multi-purpose, reconfigurable quantum circuits of unprecedented size and complexity. Control of a single phase shifter in this architecture has been used to manipulate time bin qubits \cite{honjo-ol-23-2797} and up to four photons in two spatial modes\cite{matth-np-3-346}, however, the large-scale reconfigurability required to generate arbitrary multi-photon states, including mixture and entanglement, has so far been out of reach.

Here we report an integrated quantum photonic device comprised of a two-qubit entangling gate, several Hadamard-like gates, and eight variable phase shifters, and demonstrate that it can be reconfigured with high fidelity across the complete space of possible configurations. We use this device to generate all four Bell states and perform quantum state tomography on them, to realise a Bell inequality ``manifold"---obtained from a continuum of measurement settings and states with a variable amount of entanglement---and to prepare and measure arbitrary single-photon mixed states.

\begin{figure}[t]
\centering
\includegraphics[width=3.4in]{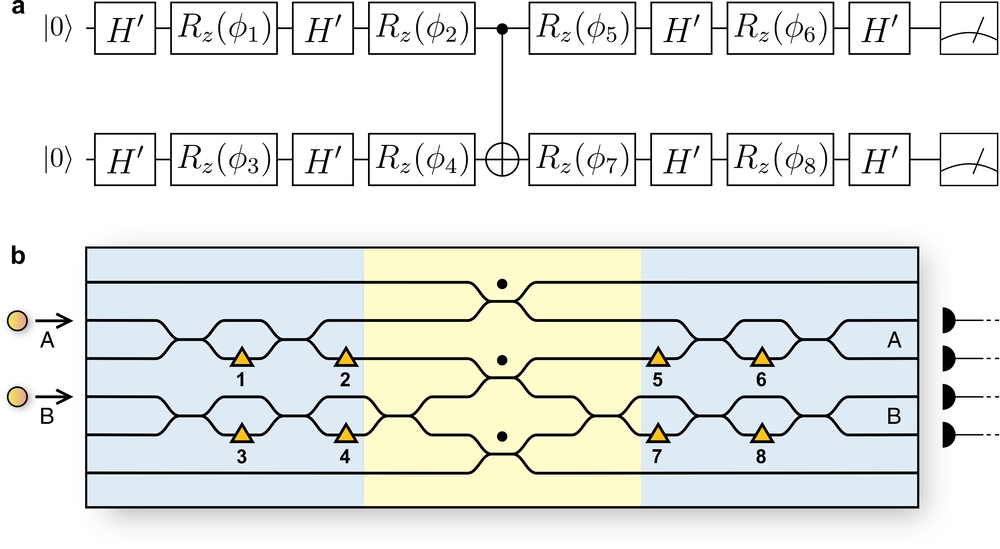}
\caption{\textbf{A two-photon reconfigurable quantum circuit for generating, manipulating and detecting entanglement and mixture.} \textbf{a} Quantum circuit diagram consisting of pairs of Hadamard-like gates 
$H'=e^{i\pi/2} e^{-i\pi\sigma_{Z}/4} H e^{-i\pi\sigma_{Z}/4}$ (where $H$ is the usual Hadamard gate)
and $R_{z}(\phi)=e^{-i\phi\sigma_{z}/2}$ rotations, that together implement $\hat{U}_{i,f}({\phi_j,\phi_k})$, and two $H'$ gates and a controlled-\textsc{sign} or \textsc{cz} gate, that together implement a \textsc{cnot}. \textbf{b} Waveguide implementation of the circuit composed of directional couplers and voltage controlled thermo-optic phase shifters drawn as orange triangles. Directional couplers with splitting ratio $\eta=1/3$ are marked with a dot, all other couplers have  $\eta=1/2$.}\label{fig:chip}\end{figure}
\vspace{10pt}
\noindent\textbf{A reconfigurable quantum photonic circuit}

\noindent The device described here is a silica-on-silicon entangling circuit, shown in Fig. \ref{fig:chip}. Two photonic qubits $A$ and $B$ are encoded in pairs of waveguides---path or dual rail encoding. These two qubits are input in the logical zero state $\ket{0_{A}} \kron \ket{0_{B}}$---
\emph{i.e.} a single photon in each upper waveguide---and are then acted upon by the quantum circuit shown in Fig.~\ref{fig:chip}.

\begin{figure*}[t]
\centering
\includegraphics[width=6.7in]{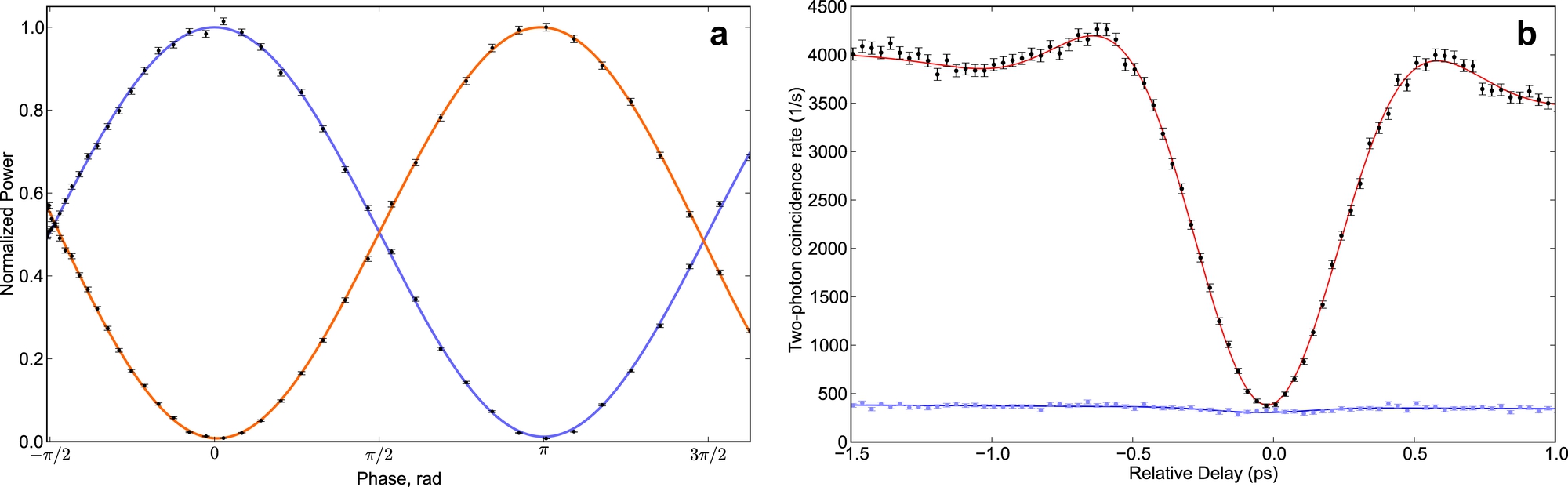}
\caption{\small \textbf{Classical and quantum interference fringes.} \textbf{a} Interference fringe measured at the two outputs of a single Mach-Zehnder (MZ) interferometer on the chip. Experimental data are presented as black circles. Solid lines show fits.   \textbf{b} Hong-Ou-Mandel dip, measured using a single MZ interferometer as a beamsplitter. Two-photon coincidence counts are shown as black circles. The red line shows a fit to this data with Gaussian and sinc components, due to quantum interference and determined by the spectral filters used, and a linear term accounting for slight decoupling of the source. The blue line shows a fit to the measured rate of accidental coincidences, with Gaussian and linear components. Error bars in both figures assume Poissonian statistics.}
\label{fig:dip}
\end{figure*}

The first part of this circuit enables arbitrary state preparation of each qubit. The central part of the circuit implements a maximally entangling postselected controlled-\textsc{not} (\textsc{cnot}) logic gate \cite{niels-a}---the canonical two-qubit entangling gate. The \textsc{cnot} gate is a postselected linear optical gate that works with probability 1/9 \cite{ralph-pra-65-62324, hofman-pra-2-024308}. The final stage of the circuit is the mirror image of the first stage and is followed by measurement in the computational basis, which together enables projective measurement of each qubit in an arbitrary basis.

The initial and final stages of the device can be reconfigured, and are each implemented using two MZ interferometers, each composed of two voltage-controlled thermal phase shifters and two directional couplers \cite{matth-np-3-346}. 
This architecture allows reconfigurable single-qubit unitary operations to be performed:
In general any unitary in $SU(2)$ can be realised using three phase shifters and an MZ interferometer \cite{reck-prl-73-58} as
$
\hat{U}_{arb}(\varphi_a, \varphi_b, \varphi_c)=e^{i\varphi_c\sigma_{z}/2}e^{i\varphi_b\sigma_{y}/2}e^{i\varphi_a\sigma_{z}/2}
$.
Here we use two phase shifters per MZ to realise 
$\hat{U}_{i}(\varphi_b,\varphi_c)=e^{-i\varphi_c\sigma_{z}/2}e^{-i\varphi_b\sigma_{y}/2}$ on each of the two input qubits, and $\hat{U}_{f}=\hat{U}_{i}^{\dagger}$ on each output qubit of the \textsc{cnot} gate. This is adequate for arbitrary, separable two-qubit state preparation and measurement, up to a global phase.
Explicitly, the entire circuit shown in Fig.~\ref{fig:chip} implements the unitary matrix
\begin{eqnarray*}
\left [ 
\hat{U}_{f}\left(\phi_{5}, \phi_{6} \right) 
\kron 
\hat{U}_{f}\left(\phi_{7}, \phi_{8} \right)
\right ] \cdot
\hat{U}_{CNOT} \cdot \\
\left [ 
\hat{U}_{i}\left(\phi_{1}, \phi_{2} \right)
\kron
\hat{U}_{i}\left(\phi_{3}, \phi_{4} \right)
\right ]
\end{eqnarray*}
where $\hat{U}_{CNOT}=\ket{00}\bra{00}+\ket{01}\bra{01}+\ket{11}\bra{10}+\ket{10}\bra{11}$ and $\phi_{1-8}$ are set by the external control voltages.

\vspace{10pt}
\noindent\textbf{Benchmarking of reconfigurability}

\noindent This circuit can be reconfigured to perform a number of different tasks, including arbitrary two-qubit pure (entangled) state preparation, arbitrary one-qubit (mixed) state preparation, state tomography, process tomography, \emph{etc.},
as detailed below. In order to characterize the precision and accuracy with which the device can be reconfigured we injected single photons into the device via a polarization maintaining optical fibre array, and measured interference fringes across each of the eight phase shifters on the chip, finding an average contrast C = 0.988 $\pm$ 0.008. (See Methods for details.) An example of a pair of such fringes is shown in Fig. \ref{fig:dip}\textbf{a}. From these measurements, we estimate the average accuracy in phase across all eight heaters to be  $\delta_{\phi} \sim 0.05$ radians (see Supplementary Information).

In addition to high-fidelity classical interference, as demonstrated in Fig. \ref{fig:dip}a, the \textsc{cnot} gate in the middle of the circuit shown in Fig \ref{fig:chip} relies on high-fidelity quantum interference\cite{laing-apl-97-211109}. Fig. \ref{fig:dip}b shows a Hong-Ou-Mandel dip\cite{hong-prl-59-2044} 
measured at a single MZ interferometer (that containing $\phi_1$) on the chip. We produced degenerate photon pairs, sharing the same spectral and polarization mode, via type-I spontaneous parametric downconversion\cite{burnh-prl-25-84} (see Methods) which were injected into the chip as shown in Fig. \ref{fig:chip}.  
 The phase in the interferometer was then set to $\pi/2$, rendering it equivalent to a $1/2$ reflectivity beamsplitter, and the two-photon coincidence count $N$ across the outputs of the interferometer was measured as a function of an off-chip optical delay between the arrival times of the two photons. The visibility of the dip $V = (N_{classical}-N_{quantum})/N_{classical}$ was measured to be $0.978\pm0.007$, taking into account the measured rate of accidental coincidences\cite{natar-apl-96-211101}. Birefringence, mode mismatch, and other such imperfections in the circuit could limit the visibility of this dip. The high visibility of the HOM dip therefore indicates the high quality of the device.

\begin{figure}[t]

\centering
\includegraphics[width=3.4in]{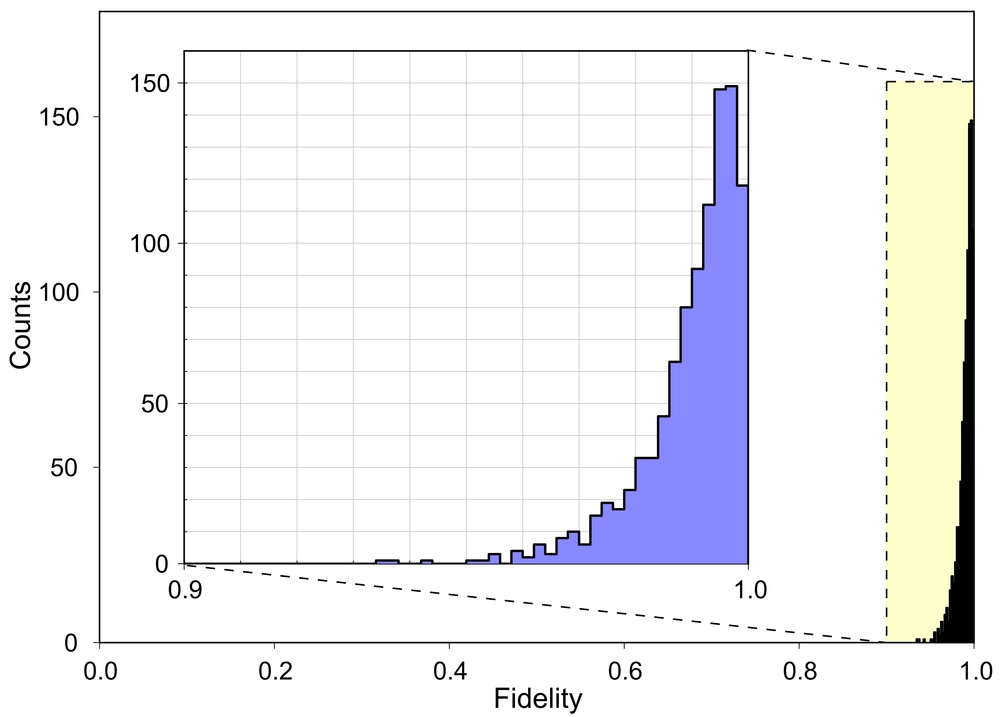}
\caption{\small \textbf{Statistical fidelity of photon .} The histogram shows the distribution of  statistical fidelity between ideal and measured coincidence probabilities, over 995 sets of eight randomly selected phases $\tilde{\varphi}$. 96\% of phase settings produced statistics corresponding with theory to $f >$ 0.97.\vspace{-0.1in}}
\label{fig:histogram}
\end{figure}

Having observed high-fidelity classical and quantum interference at individual MZ interferometers on the chip, we then used a stochastic method to characterise the operational performance of the quantum circuit as a whole, across the full space of possible configurations.
We chose, at random, $995$ vectors $\tilde{\varphi_{j}}$ representing possible configurations of the device
\begin{equation}
\tilde{\varphi_{j}} = \left [ \phi_{1}^{j}, \phi_{2}^{j},...,\phi_{8}^{j}\right]
\end{equation}
with $0\le\phi_{i}^{j}\le2\pi$. 
Injecting photon pairs as before, the probability-theoretic fidelity $f=\sum_{k}\sqrt{p_{k}\cdot p_{k}'}$
between experimentally measured coincidence probabilities at the output of the device $\left (p_{00},~p_{01},~p_{10},~p_{11} \right)$ and the ideal theoretical values $\left (p'_{00},~p'_{01},~p'_{10},~p'_{11} \right )$ was calculated for each $\tilde{\varphi_{j}}$. The statistical distribution of these fidelities is shown in Fig. \ref{fig:histogram}a. The average fidelity across $995$ configurations (equivalent to many truth tables in many bases) was measured to be 0.990$\pm$0.009 with 96\% of configurations $\tilde{\varphi_{j}}$ producing photon statistics with $f >$ 0.97. This result depends on simultaneous high fidelity quantum and classical interference, as well as accurate and precise joint control of all eight phase controllers. Poor performance of any of these component parts would result in lower fidelity output for some subset of $\{\tilde{\varphi_{j}}\}$. 
The high fidelity operation observed in these tests of reconfigurability bodes well for the operations described below.

\vspace{10pt}\noindent\textbf{Generating and characterising entanglement}

\noindent Entangled states of quantum systems are the fundamental resource in quantum information and represent the most nonclassical implication of the formalism of quantum mechanics. The circuit shown in Fig. \ref{fig:chip} can be used to prepare a continuum of entangled, partially entangled, and separable states with only computational-basis product states as input.

In order to demonstrate this ability, we first prepared and analysed each of the four maximally entangled Bell states. Inputting the $\ket{0_A}\ket{0_B}$ state as before, the state preparation stage of the circuit was used to generate each of the superposition states $\ket{\pm_A}\ket{0_B}$, $\ket{\pm_A}\ket{1_B}$, where $\ket{\pm}\equiv \left(\ket{0}\pm \ket{1} \right ) / \sqrt{2}$ , at the input of the \textsc{cnot} gate. The corresponding Bell states ($\ket{\Phi^{\pm}}$ and  $\ket{\Psi^{\pm}}$ respectively) are ideally produced by the \textsc{cnot} gate.

\begin{figure}[t!]
\centering
\includegraphics[width=2.9in]{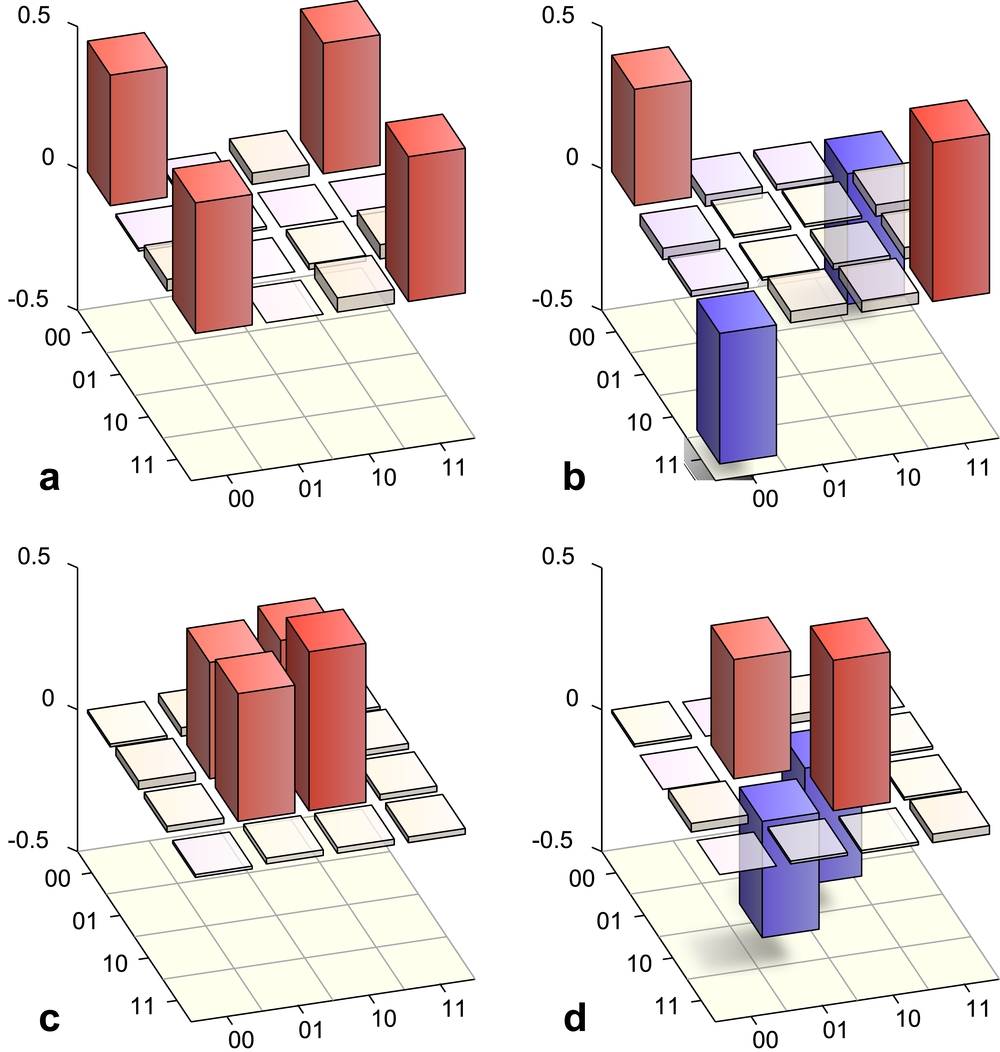}
\caption{\small \textbf{Bell states generated and characterized on-chip.} Real parts of the density operators of the states $\ket{\Phi^{+}}$, $\ket{\Phi^{-}}$, $\ket{\Psi^{+}}$ and $\ket{\Psi^{-}}$ (\textbf{a},\textbf{b},\textbf{c},\textbf{d} respectively.) \vspace{-0.1in}}

\label{fig:bells}
\end{figure}

We used the arbitrary single-qubit measurement capability of the circuit to perform maximum-likelihood quantum state tomography (QST)\cite{james-pra-64-52312} on these four states: phase shifters $\phi_{5-8}$ were used to implement each of the 16 measurements necessary to reconstruct the density operator of the state. The measured density matrices of all four Bell states are shown in Fig. \ref{fig:bells}, with quantum state fidelities $F=\left (Tr \sqrt{\sqrt{\rho_{th}} \rho_{exp} \sqrt{\rho_{th}} } \right)^{2}$ of  $0.947\pm0.002$, $0.945\pm0.002$, $0.933\pm0.002$, and $0.885\pm0.002$ respectively.

\begin{figure*}[t!]
\centering
\includegraphics[width=6.0in]{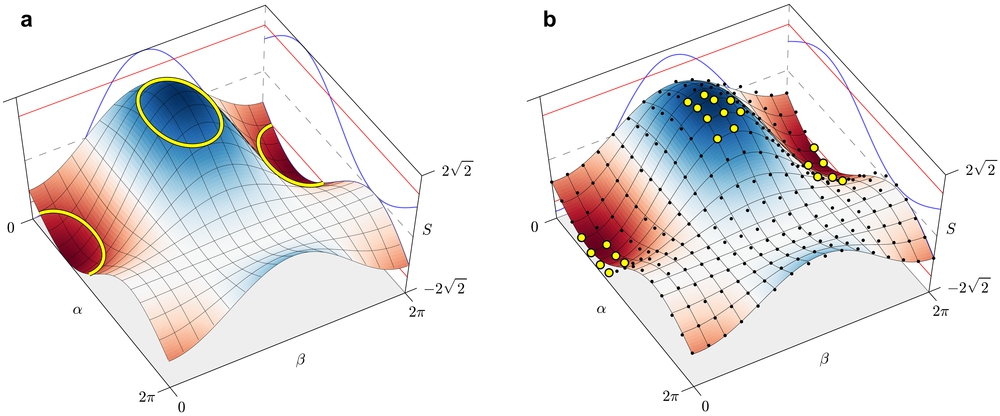}
\caption{\textbf{CHSH manifold.}
\textbf{a} The Bell-CHSH sum $S$, plotted as a function of phases $\alpha$ and $\beta$. In the $\alpha$ axis, the state shared between Alice and Bob is tuned continuously between product states at $\alpha=0, \pi$ and maximally entangled states at $\alpha=\pi/2, 3\pi/2$. The $\beta$ axis shows $S$ as a function of Bob's variable measurements, which can be thought of as two operator-axes in the real plane of the Bloch sphere, fixed with respect to each other at an angle of $\pi/2$ but otherwise free to rotate with angle $\beta$ between 0 and $2\pi$. The blue curves show a projection of the manifold onto each axis. Yellow contours mark the edges of regions of the manifold which violate $-2 \le S\le 2$. Red lines on the axes also show this limit.
\textbf{b} Experimentally measured manifold. Data points are drawn as black circles. Data points which violate the CHSH inequality are drawn as yellow circles. The surface shows a fit to the experimental data.}
\label{fig:chsh}
\end{figure*}

\vspace{10pt}\noindent\textbf{A Bell-type inequality manifold}

\noindent
The Clauser, Horne, Shimony, and Holt (CHSH)\cite{claus-prl-23-880} test of local hidden-variable models of quantum mechanics requires that the sum
\begin{equation}
S=\langle \hat{A_1} \hat{B_1} \rangle+\langle \hat{A_1} \hat{B_2} \rangle+\langle \hat{A_2} \hat{B_1} \rangle-\langle \hat{A_2} \hat{B_2} \rangle
\end{equation}
satisfies the Bell-CHSH inequality $-2 \le S \le 2$ for any local hidden-variable model, where $\hat{A}_i$, $\hat{B}_i$ are measurement operators chosen by two observers, Alice and Bob.

The Bell-CHSH experiment provides a well-known test for the presence of entanglement that we use here to examine the performance of the device, as it is reconfigured across a large parameter space. Specifically, we use $\phi_{1-4}$ and the \textsc{cnot} gate to prepare the state
\begin{equation}
\ket{\psi_{out}}=\frac{1}{2\sqrt{2}} \left [ \left ( 1-e^{i\alpha} \right ) \ket{00} + \left(1+e^{i\alpha} \right)\ket{11} \right ],
\end{equation}
where $\alpha=\phi_{1}$. By changing $\alpha$ it is thus possible to tune continuously between two orthogonal maximally entangled states: for $\alpha=0,\pi$, $\ket{\psi_{out}}$ is a product state, and with $\alpha=\pi/2,3\pi/2$, $\ket{\psi_{out}}$ is the maximally entangled state $\frac{1}{\sqrt{2}} \left ( \ket{00} \pm i \ket{11} \right )$ (up to a global phase). In the course of this preparation, we pass through a continuum of partially entangled states.
In order to evaluate $S$ we make four two-qubit measurements on the state emerging from the \textsc{cnot} gate, which correspond to combinations of observables chosen by Alice and Bob. While Alice's two measurement settings, $\phi_6 = \pi / 4, -\pi/4$, do not change, Bob's two measurement settings are varied continuously, as $\phi_8= \beta, \beta+\pi/2$. We measured $S(\alpha, \beta)$ for $\alpha \in [0,2\pi]$ and $\beta \in [0, 2\pi]$, with step size $2\pi/15$, producing the ``Bell manifold'' shown in Fig. \ref{fig:chsh}. We measured maximum and minimum values of $S$ of $2.49\pm0.03$ and $-2.54\pm0.03$ respectively. Errors were again determined by a Monte-Carlo technique, assuming Poissonian statistics. 

In order to quantitatively compare the theoretical manifold with experimental data, we used the quantity \begin{equation}R^{2}=1-\frac{\sum_{i} (S_{i}-T_{i})^{2}}{\sum_{i} (S_{i}-\bar{S})^{2}},\end{equation} where $S_{i}$ are experimentally measured values of the Bell-CHSH sum, $\bar{S}$ is the average over $S_{i}$, and $T_{i}$ are the theoretical values of $S$ shown in Fig. \ref{fig:chsh}a. In the ideal case, $R^2=1$. For the data shown in Fig. \ref{fig:chsh}b, $R^2=0.935$.

\vspace{10pt}\noindent\textbf{Generating and characterising mixture}

\noindent Mixture is often associated with noise or decoherence in quantum processes and its deliberate and controlled implementation is critical for characterisation of devices; furthermore it has been shown that quantum computing can be performed despite mixture \cite{lanyo-prl-101-200501}. More significantly, recent work has suggested that decoherence may play an important role in biological processes that exhibit quantum coherence \cite{mohse-jcp-17-174106, pleni-njop-10-113019}; photonic waveguide systems show great promise for simulating these processes \cite{perets-prl-17-170506,peruz-s-329-1500}, however, such simulations will require the controlled introduction of mixture. 

By tracing over one of the two output photons, our device can prepare an arbitrary state of a single qubit including any amount of mixture\cite{kwiat-prl-94}. The amount of entanglement in the two qubit state prepared by the first stage of the device determines the degree of mixture, which can range from zero (a pure state) to one (a maximally mixed state).

In general the state
\begin{equation}
\ket{\psi}_{out}=\alpha \gamma \ket{0_A 0_B} + \alpha \delta \ket{0_A 1_B} + \beta \gamma \ket{1_A 1_B} + \beta \delta \ket{1_A 0_B}
\end{equation}
is generated after the \textsc{cnot} gate in the circuit, where $\alpha$, $\beta$, $\gamma$, $\delta$ are complex parameters related to 
$\phi_{1-4}$. Tracing out the second qubit, we find the reduced density operator of qubit A,
\begin{eqnarray}
\rho_{A} & = & |\alpha|^{2}\ket{0}\bra{0} + \alpha \beta^* (\gamma \delta^*+\delta \gamma^*) \ket{0}\bra{1}\\
~ & + & \beta \alpha^* (\gamma \delta^*+\delta \gamma^*) \ket{1}\bra{0} \nonumber +|\beta|^{2}\ket{1}\bra{1}.
\end{eqnarray}
By choosing $\alpha$, $\beta$, $\gamma$, $\delta$, via setting $\phi_{1-4}$, the amount of mixture in this reduced density matrix can be continuously varied between 0 and 1.

\begin{figure}[t!]
\centering
\includegraphics[width=3.4in]{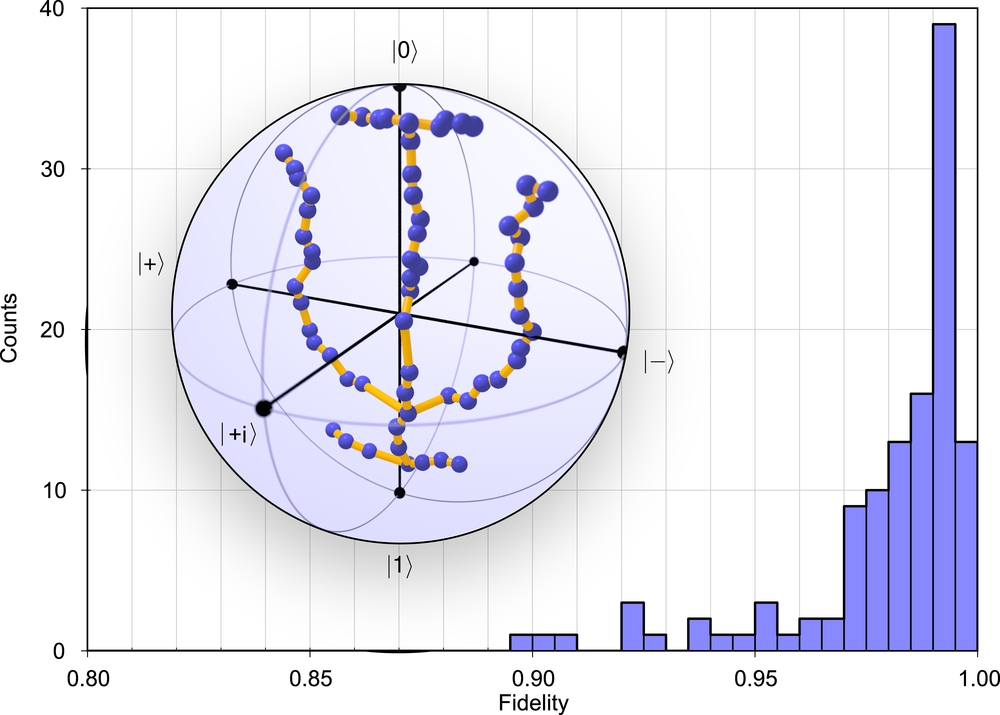}
\caption{Histogram showing the statistical distribution of quantum state fidelity between 119 randomly chosen single-qubit target states and the corresponding mixed states generated and characterized on-chip. Inset: $\Psi$ drawn in the Bloch sphere using 63 mixed states, again generated and characterized on-chip. These states are chosen from the real plane of the sphere for clarity. Note that each point is derived from a different bipartite partially entangled state.}
\label{fig:mixedstates}
\end{figure}

We chose 119 target states with various amounts of mixture, at random by the Hilbert-Schmidt measure\cite{zyczk-jopamag-34-7111}, then generated each state and reconstructed its density matrix by maximum likelihood state tomography using phase shifters $\phi_{7}$ and $\phi_{8}$. Fig. \ref{fig:mixedstates} shows the fidelity of these reconstructed states. The average quantum state fidelity across all 119 states was measured to be $0.98\pm0.02$, with 91\% of states having fidelity $> 0.95 $. We then chose 63 specific mixed states that mapped out the symbol `$\Psi$' inside the Bloch sphere, and generated them with high fidelity (Fig. \ref{fig:mixedstates}, inset).

We note that the device shown in Fig. 1 could be used to generate mixture by applying random voltages to phase shifters on a single qubit, without the need for entanglement. We chose to use the entanglement approach as a more demanding test of our device, demonstrating sufficient control to obtain the data shown in Fig. 6. An advantage of using the entanglement approach in practical applications is that it does not require pseudo-/quantum-random number generators.

\vspace{10pt}\noindent\textbf{Discussion}\\
\noindent Quantum information science and technology with photons will require circuits that are complex, stable and highly reconfigurable in a straightforward manner. 
High fidelity production and measurement of states of arbitrary entanglement and mixture will be essential for characterisation of quantum devices, and will provide a reliable means to test the unique properties of quantum physics.
The generation of mixed states may also be important in quantum photonic analogues of biochemical systems that rely on decoherence \cite{mohse-jcp-17-174106, pleni-njop-10-113019}. 

Although on-chip polarization encoding is possible\cite{sanso-prl-105-200503}, the inherent interferometric stability of integrated optics makes path encoding of qubits a natural choice, with the further advantage that encoding of higher-dimensional qudits\cite{rossi-prl-102-153902} is immediately possible. This is in contrast with bulk optics, where two-level polarization encoding is more natural, and stable path encoding requires a considerable resource overhead. Furthermore, this architecture could be used to manipulate hyper-entanglement \cite{ceccar-prl-103-160401} encoded with multiple degrees of freedom \cite{saleh-oe-18-20475, saleh-pjieee-2-736}.
Circuits such as the one presented here could be used in conjunction with adaptive (classical) algorithms to bypass the need for calibration of the phase shifters in particular applications. For example, repeated measurement and feedback onto the voltage-controlled phase shifters based on comparison of the output state with a desired target state could be used to reconfigure the circuit via a genetic algorithm.

\vspace{10pt}\noindent\textbf{Methods} \\
\noindent \textit{Device:}
The waveguide device was fabricated on a silicon wafer, upon which a 16$\mu$m layer of undoped silica was deposited to form the lower cladding of the waveguides. 3.5-$\mu$m-wide waveguides were then patterned in a 3.5-$\mu$m layer of silica doped with germanium and boron oxides. A 16-$\mu$m layer of silica, doped with phosphorous and boron so as to be index-matched with the lower layer, constitutes the upper cladding. Resistive heaters and corresponding electric contacts were then patterned in metal on top of the chip using standard lithographic techniques. Dimensions of the chip are 70mm $\times$ 3mm.

\vspace{0.05in}
\noindent\textit{Photon Source:}
Degenerate pairs of 808 nm photons were generated by focusing a 404 nm, 60mW laser onto a 2mm thick, Bisumuth Borate $\textrm{BiB}_4\textrm{O}_6$ (BiBO) nonlinear crystal, phase matched for type I spontaneous parametric down conversion (opening angle of 3 degrees). 

Photon pairs were spectrally filtered using 3nm full-width at half maximum interference filters. The interference filters were designed with central wavelength 808 nm and were tilted to ensure that photon pairs were identically filtered. Photons were then collected into polarisation maintaining fibre (PMF) using 11mm aspheric lenses. Typical two-photon coincidence count rates of 100kHz  were achieved using $\sim60\%$ efficient, silicon based avalanche photo-diode single photon counting modules (SPCM). The photons were then launched into the waveguide chip by butt-coupling arrays of PMF with 250$\mu$m spacing (matched to the input and output waveguide pitch). Photons were collected from the output of the chip also using arrays of polarisation maintaining fibre  and detected with fibre coupled SPCMs. A typical facet-to-facet coupling efficiency of $\sim60\%$ was achieved.

\vspace{0.05in}
\noindent \textit{Calibration of Phase Shifters:}
Each thermal phase shifter $\phi_{i}$ has a nonlinear phase voltage relationship:
\begin{equation}
\phi_{i}(V_{i}) = \alpha_{i} + \beta_{i} V^{2} + \gamma_{i} V^{3} + \delta_{i} V^{4},
\label{eqn:phasevoltage}
\end{equation}
where $V_{i}$ is the voltage applied across phase shifter $i$ and $\phi_{i}$ is the resulting phase shift. $\alpha_{i}$, $\beta_{i}$, $\gamma_{i}$ and $\delta_{i}$ are real numbers associated with the response of a particular heater. Each phase shifter can be seen to occupy one particular MZ interferometer in the circuit. $\phi_{2}$ and $\phi_{5}$ can be seen as acting on a single, lossy MZ interferometer.

Each phase shifter in the circuit was calibrated as follows: Bright light from an 810nm 1.3mW laser was injected into one input port of each MZ, and the intensity at each output port was measured as a function of the voltage applied across the heater, which was swept linearly between 0V and 7V. This produced classical interference fringes, distorted by the nonlinear phase-voltage relationship (\ref{eqn:phasevoltage}). We then fitted the function 
\begin{equation}
I(V)=A(1-C \cdot cos^{2}(\phi(V)/2))
\end{equation}
to each set of experimental data with $A$, $C$, $\alpha$, $\beta$, $\gamma$ and $\delta$ as fitting parameters. This yields the complete (approximate) phase-voltage relationship for each heater. No evidence of crosstalk between phase shifters (due to thermal effects or otherwise) was observed in these experiments.

\vspace{0.05in}
\noindent \textit{Single-Photon Fringes:}
We measured single photon fringes for each phase shifter using photon pairs from the source. Injecting photons from one arm of the source into the chip, we counted coincidences between single photon events from a particular output of the interferometer, and those from the other arm of the source. This approach largely mitigates the contribution of SPCM dark counts. The fit shown in Fig. \ref{fig:dip} has the form $A(1-C \cdot cos^{2}(\phi/2))$ where $C$ is the fringe contrast and $A$ is the amplitude. Each fringe is normalized with respect to its amplitude for clarity.

\vspace{-5pt}
\section*{Acknowledgements}\label{sec:acknowledgements}\vspace{-5pt}
{\small We thank N. Brunner, J. G. Rarity and P. Ivanov for helpful contributions. This work was supported by the Engineering and Physical Sciences Research Council (EPSRC), the European Research Council (ERC), Intelligence Advanced Research Projects Activity (IARPA), the Leverhulme Trust, the Centre for Nanoscience and Quantum Information (NSQI), PHORBITECH, the Quantum Information Processing Interdisciplinary Research Collaboration (QIP IRC), and the Quantum Integrated Photonics (QUANTIP) project. J.L.O'B. acknowledges a Royal Society Wolfson Merit Award. M.L. acknowledges the Marie Curie International  Incoming Fellowship.}

\section*{Author Contributions}\label{sec:contributions}
All authors contributed extensively to the work presented in this paper.

\section*{Additional information}\label{sec:info}
The authors declare no competing financial interests. Supplementary information accompanies this paper at www.nature.com/naturephotonics. Reprints and permission information is available online at http://www.nature.com/reprints. Correspondence and requests for materials should be addressed to J.L.O'B.


\begin{thebibliography}{10}
\expandafter\ifx\csname url\endcsname\relax
  \def\url#1{\texttt{#1}}\fi
\expandafter\ifx\csname urlprefix\endcsname\relax\def\urlprefix{URL }\fi
\providecommand{\bibinfo}[2]{#2}
\providecommand{\eprint}[2][]{\url{#2}}

\bibitem{niels-a}
\bibinfo{author}{Nielsen, M.~A.} \& \bibinfo{author}{Chuang, I.~L.}
\newblock \emph{\bibinfo{title}{{Quantum Computation and Quantum Information}}}
  (\bibinfo{publisher}{Cambridge University Press}, \bibinfo{year}{2000}).

\bibitem{deuts-potrsolamaps-400-97}
\bibinfo{author}{Deutsch, D.}
\newblock \bibinfo{title}{{Quantum Theory, the Church-Turing Principle and the
  Universal Quantum Computer}}.
\newblock \emph{\bibinfo{journal}{Proc. Roy. Soc. Lond. A}}
  \textbf{\bibinfo{volume}{400}}, \bibinfo{pages}{97--117}
  (\bibinfo{year}{1985}).

\bibitem{feynm-ijotp-21-467}
\bibinfo{author}{Feynman, R.~P.}
\newblock \bibinfo{title}{{Simulating Physics with Computers}}.
\newblock \emph{\bibinfo{journal}{International Journal of Theoretical
  Physics}} \textbf{\bibinfo{volume}{21}}, \bibinfo{pages}{467--488}
  (\bibinfo{year}{1982}).

\bibitem{lloyd-s-273-1073}
\bibinfo{author}{Lloyd, S.}
\newblock \bibinfo{title}{{Universal Quantum Simulators}}.
\newblock \emph{\bibinfo{journal}{Science}} \textbf{\bibinfo{volume}{273}},
  \bibinfo{pages}{1073--1078} (\bibinfo{year}{1996}).

\bibitem{gisin-np-1-165}
\bibinfo{author}{Gisin, N.} \& \bibinfo{author}{Thew, R.}
\newblock \bibinfo{title}{{Quantum communication}}.
\newblock \emph{\bibinfo{journal}{Nature Photon.}}
  \textbf{\bibinfo{volume}{1}}, \bibinfo{pages}{165--171}
  (\bibinfo{year}{2007}).

\bibitem{giova-s-306-1330}
\bibinfo{author}{Giovannetti, V.}, \bibinfo{author}{Lloyd, S.} \&
  \bibinfo{author}{Maccone, L.}
\newblock \bibinfo{title}{{Quantum-Enhanced Measurements: Beating the Standard
  Quantum Limit}}.
\newblock \emph{\bibinfo{journal}{Science}} \textbf{\bibinfo{volume}{306}},
  \bibinfo{pages}{1330--1336} (\bibinfo{year}{2004}).
\newblock.

\bibitem{dowli-cp-49-125}
\bibinfo{author}{Dowling, J.~P.}
\newblock \bibinfo{title}{{Quantum optical metrology---the lowdown on high-N00N
  states}}.
\newblock \emph{\bibinfo{journal}{Contemp. Phys.}}
  \textbf{\bibinfo{volume}{49}}, \bibinfo{pages}{125--143}
  (\bibinfo{year}{2008}).

\bibitem{o'bri-np-3-687}
\bibinfo{author}{O'Brien, J.~L.}, \bibinfo{author}{Furusawa, A.} \&
  \bibinfo{author}{Vuckovic, J.}
\newblock \bibinfo{title}{{Photonic quantum technologies}}.
\newblock \emph{\bibinfo{journal}{Nature Photon.}}
  \textbf{\bibinfo{volume}{3}}, \bibinfo{pages}{687--695}
  (\bibinfo{year}{2009}).

\bibitem{clark-pra-79-30303}
\bibinfo{author}{Clark, A.~S.}, \bibinfo{author}{Fulconis, J.},
  \bibinfo{author}{Rarity, J.~G.}, \bibinfo{author}{Wadsworth, W.~J.} \&
  \bibinfo{author}{O'Brien, J.~L.}
\newblock \bibinfo{title}{{All-optical-fiber polarization-based quantum logic
  gate}}.
\newblock \emph{\bibinfo{journal}{Physical Review A}}
  \textbf{\bibinfo{volume}{79}}, \bibinfo{pages}{030303+}
  (\bibinfo{year}{2009}).

\bibitem{hall-prl-106-53901}
\bibinfo{author}{Hall, M.~A.}, \bibinfo{author}{Altepeter, J.~B.} \&
  \bibinfo{author}{Kumar, P.}
\newblock \bibinfo{title}{Ultrafast switching of photonic entanglement}.
\newblock \emph{\bibinfo{journal}{Physical Review Letters}}
  \textbf{\bibinfo{volume}{106}}, \bibinfo{pages}{053901+}
  (\bibinfo{year}{2011}).

\bibitem{polit-s-320-646}
\bibinfo{author}{Politi, A.}, \bibinfo{author}{Cryan, M.~J.},
  \bibinfo{author}{Rarity, J.~G.}, \bibinfo{author}{Yu, S.} \&
  \bibinfo{author}{O'Brien, J.~L.}
\newblock \bibinfo{title}{{Silica-on-Silicon Waveguide Quantum Circuits}}.
\newblock \emph{\bibinfo{journal}{Science}} \textbf{\bibinfo{volume}{320}},
  \bibinfo{pages}{646--649} (\bibinfo{year}{2008}).

\bibitem{matth-np-3-346}
\bibinfo{author}{Matthews, J. C.~F.}, \bibinfo{author}{Politi, A.},
  \bibinfo{author}{Stefanov, A.} \& \bibinfo{author}{O'Brien, J.~L.}
\newblock \bibinfo{title}{{Manipulation of multiphoton entanglement in
  waveguide quantum circuits}}.
\newblock \emph{\bibinfo{journal}{Nature Photon.}}
  \textbf{\bibinfo{volume}{3}}, \bibinfo{pages}{346--350}
  (\bibinfo{year}{2009}).

\bibitem{smith-oe17-13516}
\bibinfo{author}{Smith, B.~J.}, \bibinfo{author}{Kundys, D.},
  \bibinfo{author}{Thomas-Peter, N.}, \bibinfo{author}{Smith, P. G.~R.} \&
  \bibinfo{author}{Walmsley, I.~A.}
\newblock \bibinfo{title}{Phase-controlled integrated photonic quantum
  circuits}.
\newblock \emph{\bibinfo{journal}{Opt. Express}} \textbf{\bibinfo{volume}{17}},
  \bibinfo{pages}{13516--13525} (\bibinfo{year}{2009}).
\newblock.

\bibitem{polit-s-325-1221}
\bibinfo{author}{Politi, A.}, \bibinfo{author}{Matthews, J. C.~F.} \&
  \bibinfo{author}{O'Brien, J.~L.}
\newblock \bibinfo{title}{{Shor's Quantum Factoring Algorithm on a Photonic
  Chip}}.
\newblock \emph{\bibinfo{journal}{Science}} \textbf{\bibinfo{volume}{325}},
  \bibinfo{pages}{1221+} (\bibinfo{year}{2009}).

\bibitem{sanso-prl-105-200503}
\bibinfo{author}{Sansoni, L.} \emph{et~al.}
\newblock \bibinfo{title}{{Polarization Entangled State Measurement on a
  Chip}}.
\newblock \emph{\bibinfo{journal}{Physical Review Letters}}
  \textbf{\bibinfo{volume}{105}}, \bibinfo{pages}{200503+}
  (\bibinfo{year}{2010}).

\bibitem{laing-apl-97-211109}
\bibinfo{author}{Laing, A.} \emph{et~al.}
\newblock \bibinfo{title}{{High-fidelity operation of quantum photonic
  circuits}}.
\newblock \emph{\bibinfo{journal}{Appl. Phys. Lett.}}
  \textbf{\bibinfo{volume}{97}}, \bibinfo{pages}{211109+}
  (\bibinfo{year}{2010}).

\bibitem{peruzz-nc-2-224}
\bibinfo{author}{Peruzzo, A.}, \bibinfo{author}{Laing, A.},
  \bibinfo{author}{Politi, A.}, \bibinfo{author}{Rudolph, T.} \&
  \bibinfo{author}{O'Brien, J.~L.}
\newblock \bibinfo{title}{Multimode quantum interference of photons in
  multiport integrated devices}.
\newblock \emph{\bibinfo{journal}{Nature Commun.}}
  \textbf{\bibinfo{volume}{2}}, \bibinfo{pages}{224+} (\bibinfo{year}{2011}).

\bibitem{honjo-ol-23-2797}
\bibinfo{author}{Honjo, T.}, \bibinfo{author}{Inoue, K.} \&
  \bibinfo{author}{Takahashi, H.}
\newblock \bibinfo{title}{Differential-phase-shift quantum key distribution
  experiment with aplanar light-wave circuit {Mach-Zehnderinterferometer}}.
\newblock \emph{\bibinfo{journal}{Opt. Lett.}} \textbf{\bibinfo{volume}{29}},
  \bibinfo{pages}{2797--2799} (\bibinfo{year}{2004}).

\bibitem{ralph-pra-65-62324}
\bibinfo{author}{Ralph, T.~C.}, \bibinfo{author}{Langford, N.~K.},
  \bibinfo{author}{Bell, T.~B.} \& \bibinfo{author}{White, A.~G.}
\newblock \bibinfo{title}{{Linear optical controlled-NOT gate in the
  coincidence basis}}.
\newblock \emph{\bibinfo{journal}{Physical Review A}}
  \textbf{\bibinfo{volume}{65}}, \bibinfo{pages}{062324+}
  (\bibinfo{year}{2002}).

\bibitem{hofman-pra-2-024308}
\bibinfo{author}{Hofmann, H.~F.} \& \bibinfo{author}{Takeuchi, S.}
\newblock \bibinfo{title}{Quantum phase gate for photonic qubits using only
  beam splitters and postselection}.
\newblock \emph{\bibinfo{journal}{Physical Review A}}
  \textbf{\bibinfo{volume}{66}}, \bibinfo{pages}{024308+}
  (\bibinfo{year}{2002}).

\bibitem{reck-prl-73-58}
\bibinfo{author}{Reck, M.}, \bibinfo{author}{Zeilinger, A.},
  \bibinfo{author}{Bernstein, H.~J.} \& \bibinfo{author}{Bertani, P.}
\newblock \bibinfo{title}{{Experimental realization of any discrete unitary
  operator}}.
\newblock \emph{\bibinfo{journal}{Physical Review Letters}}
  \textbf{\bibinfo{volume}{73}}, \bibinfo{pages}{58--61}
  (\bibinfo{year}{1994}).

\bibitem{hong-prl-59-2044}
\bibinfo{author}{Hong, C.~K.}, \bibinfo{author}{Ou, Z.~Y.} \&
  \bibinfo{author}{Mandel, L.}
\newblock \bibinfo{title}{{Measurement of subpicosecond time intervals between
  two photons by interference}}.
\newblock \emph{\bibinfo{journal}{Physical Review Letters}}
  \textbf{\bibinfo{volume}{59}}, \bibinfo{pages}{2044--2046}
  (\bibinfo{year}{1987}).

\bibitem{burnh-prl-25-84}
\bibinfo{author}{Burnham, D.~C.} \& \bibinfo{author}{Weinberg, D.~L.}
\newblock \bibinfo{title}{{Observation of Simultaneity in Parametric Production
  of Optical Photon Pairs}}.
\newblock \emph{\bibinfo{journal}{Physical Review Letters}}
  \textbf{\bibinfo{volume}{25}}, \bibinfo{pages}{84--87}
  (\bibinfo{year}{1970}).

\bibitem{natar-apl-96-211101}
\bibinfo{author}{Natarajan, C.~M.} \emph{et~al.}
\newblock \bibinfo{title}{{Operating quantum waveguide circuits with
  superconducting single-photon detectors}}.
\newblock \emph{\bibinfo{journal}{Appl. Phys. Lett.}}
  \textbf{\bibinfo{volume}{96}}, \bibinfo{pages}{211101+}
  (\bibinfo{year}{2010}).

\bibitem{james-pra-64-52312}
\bibinfo{author}{James, D. F.~V.}, \bibinfo{author}{Kwiat, P.~G.},
  \bibinfo{author}{Munro, W.~J.} \& \bibinfo{author}{White, A.~G.}
\newblock \bibinfo{title}{{Measurement of qubits}}.
\newblock \emph{\bibinfo{journal}{Physical Review A}}
  \textbf{\bibinfo{volume}{64}}, \bibinfo{pages}{052312+}
  (\bibinfo{year}{2001}).

\bibitem{claus-prl-23-880}
\bibinfo{author}{Clauser, J.~F.}, \bibinfo{author}{Horne, M.~A.},
  \bibinfo{author}{Shimony, A.} \& \bibinfo{author}{Holt, R.~A.}
\newblock \bibinfo{title}{{Proposed Experiment to Test Local Hidden-Variable
  Theories}}.
\newblock \emph{\bibinfo{journal}{Physical Review Letters}}
  \textbf{\bibinfo{volume}{23}}, \bibinfo{pages}{880--884}
  (\bibinfo{year}{1969}).

\bibitem{lanyo-prl-101-200501}
\bibinfo{author}{Lanyon, B.~P.}, \bibinfo{author}{Barbieri, M.},
  \bibinfo{author}{Almeida, M.~P.} \& \bibinfo{author}{White, A.~G.}
\newblock \bibinfo{title}{{Experimental Quantum Computing without
  Entanglement}}.
\newblock \emph{\bibinfo{journal}{Physical Review Letters}}
  \textbf{\bibinfo{volume}{101}}, \bibinfo{pages}{200501+}
  (\bibinfo{year}{2008}).

\bibitem{mohse-jcp-17-174106}
\bibinfo{author}{Mohseni, M.}, \bibinfo{author}{Rebentrost, P.},
  \bibinfo{author}{Lloyd, S.} \& \bibinfo{author}{Guzik, A.~A.}
\newblock \bibinfo{title}{Environment-assisted quantum walks in photosynthetic
  energy transfer}.
\newblock \emph{\bibinfo{journal}{The Journal of Chemical Physics}}
  \textbf{\bibinfo{volume}{129}}, \bibinfo{pages}{174106+}
  (\bibinfo{year}{2008}).
\newblock.

\bibitem{pleni-njop-10-113019}
\bibinfo{author}{Plenio, M.~B.} \& \bibinfo{author}{Huelga, S.~F.}
\newblock \bibinfo{title}{{Dephasing-assisted transport: quantum networks and
  biomolecules}}.
\newblock \emph{\bibinfo{journal}{New Journal of Physics}}
  \textbf{\bibinfo{volume}{10}}, \bibinfo{pages}{113019+}
  (\bibinfo{year}{2008}).

\bibitem{perets-prl-17-170506}
\bibinfo{author}{Perets, H.~B.} \emph{et~al.}
\newblock \bibinfo{title}{Realization of quantum walks with negligible
  decoherence in waveguide lattices}.
\newblock \emph{\bibinfo{journal}{Physical Review Letters}}
  \textbf{\bibinfo{volume}{100}}, \bibinfo{pages}{170506+}
  (\bibinfo{year}{2008}).

\bibitem{peruz-s-329-1500}
\bibinfo{author}{Peruzzo, A.} \emph{et~al.}
\newblock \bibinfo{title}{{Quantum Walks of Correlated Photons}}.
\newblock \emph{\bibinfo{journal}{Science}} \textbf{\bibinfo{volume}{329}},
  \bibinfo{pages}{1500--1503} (\bibinfo{year}{2010}).

\bibitem{kwiat-prl-94}
\bibinfo{author}{Peters, N.~A.}, \bibinfo{author}{Barreiro, J.~T.},
  \bibinfo{author}{Goggin, M.~E.}, \bibinfo{author}{Wei, T.~C.} \&
  \bibinfo{author}{Kwiat, P.~G.}
\newblock \bibinfo{title}{Remote state preparation: Arbitrary remote control of
  photon polarization}.
\newblock \emph{\bibinfo{journal}{Physical Review Letters}}
  \bibinfo{pages}{150502+} (\bibinfo{year}{2005}).

\bibitem{zyczk-jopamag-34-7111}
\bibinfo{author}{Zyczkowski, K.} \& \bibinfo{author}{Sommers, H.-J.}
\newblock \bibinfo{title}{{Induced measures in the space of mixed quantum
  states}}.
\newblock \emph{\bibinfo{journal}{J. Phys. A.}} \textbf{\bibinfo{volume}{34}},
  \bibinfo{pages}{7111--7125} (\bibinfo{year}{2001}).

\bibitem{rossi-prl-102-153902}
\bibinfo{author}{Rossi, A.}, \bibinfo{author}{Vallone, G.},
  \bibinfo{author}{Chiuri, A.}, \bibinfo{author}{De~Martini, F.} \&
  \bibinfo{author}{Mataloni, P.}
\newblock \bibinfo{title}{Multipath entanglement of two photons}.
\newblock \emph{\bibinfo{journal}{Physical Review Letters}}
  \textbf{\bibinfo{volume}{102}}, \bibinfo{pages}{153902+}
  (\bibinfo{year}{2009}).

\bibitem{ceccar-prl-103-160401}
\bibinfo{author}{Ceccarelli, R.}, \bibinfo{author}{Vallone, G.},
  \bibinfo{author}{De~Martini, F.}, \bibinfo{author}{Mataloni, P.} \&
  \bibinfo{author}{Cabello, A.}
\newblock \bibinfo{title}{Experimental entanglement and nonlocality of a
  {Two-Photon} {Six-Qubit} cluster state}.
\newblock \emph{\bibinfo{journal}{Physical Review Letters}}
  \textbf{\bibinfo{volume}{103}}, \bibinfo{pages}{160401+}
  (\bibinfo{year}{2009}).

\bibitem{saleh-oe-18-20475}
\bibinfo{author}{Saleh, M.~F.}, \bibinfo{author}{Di~Giuseppe, G.},
  \bibinfo{author}{Saleh, B. E.~A.} \& \bibinfo{author}{Teich, M.~C.}
\newblock \bibinfo{title}{Modal and polarization qubits in {Ti:LiNbO3} photonic
  circuits for a universal quantum logic gate}.
\newblock \emph{\bibinfo{journal}{Opt. Express}} \textbf{\bibinfo{volume}{18}},
  \bibinfo{pages}{20475--20490} (\bibinfo{year}{2010}).

\bibitem{saleh-pjieee-2-736}
\bibinfo{author}{Saleh, M.~F.}, \bibinfo{author}{Di~Giuseppe, G.},
  \bibinfo{author}{Saleh, B. E.~A.} \& \bibinfo{author}{Teich, M.~C.}
\newblock \bibinfo{title}{Photonic circuits for generating modal, spectral, and
  polarization entanglement}.
\newblock \emph{\bibinfo{journal}{IEEE Photon. J.}}
  \textbf{\bibinfo{volume}{2}}, \bibinfo{pages}{736--752}
  (\bibinfo{year}{2010}).

\end{thebibliography}
\end{document}